\documentclass[10pt,twocolumn,twoside]{IEEEtran}

\usepackage{amsmath,epsfig}

\newcommand{\R}{I\!\! R}

\newcommand{\Z}{Z\!\!\! Z}

\begin{document}

\title{Delay-Coordinates Embeddings as a Data Mining Tool 
for Denoising Speech Signals.}

\author{D. Napoletani$^1$, C.A. Berenstein$^2$, T. Sauer$^{3a}$, D.C. Struppa$^{3b}$ and D. Walnut$^{3c}$.
\thanks{$^1$ School of Computational Sciences, George Mason University, Fairfax, VA  22030, email:dnapolet@gmu.edu}
\thanks{$^2$ Institute for Systems Research, University of Maryland, College Park, MD  20742,email:carlos@glue.umd.edu}
\thanks{$^{3a,b,c}$ Department of Mathematical Sciences,George Mason University,Fairfax, 
VA  22030,emails: $^{3a}$ tsauer@gmu.edu; $^{3b}$ dstruppa@gmu.edu; $^{3c}$ dwalnut@gmu.edu}}

\maketitle

\begin{abstract}

In this paper we utilize techniques from the theory of 
non-linear dynamical systems to define a notion of embedding 
threshold 
estimators. More specifically we use delay-coordinates 
embeddings 
of sets of coefficients of the measured signal (in some 
chosen frame) as a data 
mining tool to separate structures that are likely to be 
generated
by signals belonging to some predetermined data set.
We describe a particular variation of the embedding
 threshold estimator
implemented in  a windowed Fourier frame, 
and we apply it to  speech signals heavily corrupted with 
the addition of 
several types of white noise. Our experimental work 
seems to suggest that, after training on the data sets of 
interest,
these estimators perform  well for a 
variety of  white noise processes and noise intensity levels.
The method is compared, for the case of Gaussian white noise, 
to a block thresholding estimator.

\end{abstract}


\noindent

\section{Introduction}

In this paper we explore the performance of a method of denoising that 
is designed to be  efficient for a variety of white noise 
contaminations and noise intensities, while keeping a fixed choice of  
parameters  of the algorithm itself (adapted to the class of signals to denoise).
The method is based on a loose distinction between 
the geometry of
delay-coordinates embeddings of, respectively, deterministic time 
series and non-deterministic 
ones.  Delay-coordinates 
embeddings are  the basis of many applications of 
the theory of non-linear
dynamical systems, see for example  [ASY] or [KS], 
our work stands apart from previous applications 
of embeddings in that  
no {\it exact} modelization of the underlyning signals (through the 
delay-coordinates embeddings) is needed nor  
attempted here. Instead, we measure the overall `squeezing' of the dynamics 
along
the principal direction of the embedding image by computing the quotient of the 
largest and smallest singular values.
\\
\\
We define first of all the context in which we look for signal estimators.
Let $F[n]$, $n=1,...,N$, be a discrete signal of length $N$,  and 
let
$X[n]=F[n]+W[n]$, $n=1,...,N$, be  a contaminated measurement of $F[n]$,
 where $W[n]$ are
realizations
of a white noise process $W$, throughout this paper we use the notation 
$E(*)$ to denote the expected value of a quantity $*$. 
\\
\\
 Generally we are interested in estimators $F$ such that
the expected mean square error $E\{|f-F|^2\}$ is as small as possible. 
For a given 
discrete orthonormal basis $B=\{g_m\}$ of the $N$ dimensional space of 
discrete signals, we can write:
$X=\sum_{m=0}^{N-1} X_B[m]g_m$ where $X_B[m]=<X,g_m>$ is the inner 
product of $X$ and $g_m$. 
Given such 
notation, we can define a class of estimators that is amenable to 
theoretical analysis, namely
the 
class of diagonal estimators of the form
$\tilde F=\sum_{m=0}^{N-1} d_m(X_B[m])g_m$ where $d_m(X_B[m])$ is a 
function that depends only 
on the value
of $X_B[m]$.
One particular kind of diagonal estimator is the hard thresholding 
estimator $\tilde F_T$
(for $T$ some positive real number) defined by the choice
\begin{equation}
\tilde F_T=\sum_{m=0}^{N-1} d_m(X_B[m])g_m
\end{equation}
where 
$$d_m(X_B[m])=X_B[m] \,\, \text{if} \,\,|X_B[m]|>T\,$$
and
$$d_m(X_B[m])=0\,\,\ \text{otherwise}.$$
\\
If $W[n]$ are realizations of a white normal distribution with variance 
$\sigma^2$, then 
it is shown in [DJ]
that $\tilde F_T$, with $T=\sigma\sqrt{2logN}$, achieves almost minimax 
risk
(when implemented in 
a wavelet basis) for the class of signals $f[n]$ of bounded variation.
The possibility of proving such a striking result is based, in part, on 
the fact that the
coefficients
$W_B[n]$ are realizations of a Gaussian white noise process in 
any basis $B$.
\\
\\
Several techniques have been developed to deal
 with the non-Gaussian case, some of the most successful are 
the Efromovich-Pinsker (EP) estimator (see for example 
[ELPT] and references threin) and the block threshold
estimators of Cai and collaborators (see [CS],[C] and the more recent [CL]).
In these methods, the variance of 
the white process needs to be estimated from the data,
moreover, since the threshold is designed to evaluate
intensities (or relative intensities) of the coefficients in blocks 
of multiwavelets, low intensity details may be filtered out 
as it is the case for simpler denoising methods (see also
remark 3 on the issue of low intensity non-noisy features).
\\
\\
The method we describe in this paper 
does not need 
the knowledge of the noise intensity level (thanks to the use
of {\it quotients} of singular values), and it is 
remarkably robust to changes in the type of noise distribution.
\\ 
This strenght is achieved at a price,
the inner parameters of the algorithm need to be adjusted to 
the data, this is true to some extent for the EP  and block thresholding
algorithms as 
well (see again [ELPT] and [CL]),
but the number and type of parameters that need to be trained 
in our 
approach is increased by the need of choosing a `good' delay-coordinates 
embedding suitable for the data we would like to denoise.
\\
In  
section V we will explore possible ways to make the training 
on the data 
automatic, but it is 
yet to be seen at this stage
which data sets are amenable to the analysis we propose. 
This paper is meant
as a mostly experimental analysis that suggest the method is 
sound at least for 
one choice of data sets (namely, speech signals).
\\
\\
Because of the choice of applying our algorithm to a database of 
speech signals, we decided to use windowed Fourier frames as a basic
analytical tool. This is an obvious way in which we are already adapting to 
the data, but more general frames $\mathcal D$ could be used, or even 
collection of frames and bases, therefore we prefer to label
$\mathcal D$ as a {\it dictionary} of analysis.  
\\
Note that any discrete periodic signal $X[n]$, $n\in \Z$ with period 
$N$ can 
be represented in a discrete 
windowed Fourier frame. The atoms in this frame are of the form
\begin{equation}
g_{m,l}[n]=g[n-m]exp(-\frac{i2 \pi ln}{N}),\,\,\,n\in \Z.
\end{equation}
We choose the window $g$ to be a symmetric $N$-periodic function of norm $1$ and 
support $q$. Specifically we can choose $g$ to be the characteristic function of the 
$[0,1]$ interval; we realize that this may not be the most robust choice
in many cases, but we have deliberately selected this function to avoid 
excessive smoothing
which was found to adversely affect our algorithm.
\\
\\ 
Under the previous conditions $x$ can be completely reconstructed from the 
inner products
$\mathcal F X[m,l]=<X,g_{m,l}>$, i.e.,
\begin{equation}
X=\frac{1}{N} \sum_{m=0}^{N-1} \sum_{l=0}^{N-1} \mathcal F X[m,l] \tilde g_{m,l}
\end{equation}
where
\begin{equation}
\tilde g_{m,l}[n]=g[n-m]\exp(\frac{i2 \pi ln}{N}),\,\,\,n\in \Z
\end{equation}
We denote the collection $\{<X,g_{m,l}>\}$ by $\mathcal F X$. 
For finite discrete signals of length $N$ the reconstruction has boundary errors. However,
the region affected by such boundary
effects is limited by the size $q$ of the support of $g$ and we can therefore have 
perfect reconstruction if we first extend $X$
suitably at the boundaries of its support and then compute the 
inner products $\mathcal F X$. 
More details 
can be found in [S] and references therein. 
\\
\\
Since for speech signals 
much of the structure in the time frequency domain is contained in  localized `ridges' 
that are oriented 
in time direction, the collection $C_p$ of double-indexed
 paths 
\begin{equation}
\gamma_{\bar m, \bar l}=\{g_{m,l} \,\, \text{such that}\,\, l=\bar l, \bar m\leq m\leq \bar m+p\},
\end{equation}
where $p$ is some positive integer, will be relatively sensitive to local time 
changes of such ridges, since each path is a short line in the time frequency
domain oriented in the time direction. 
\\
\\ 
The choice of $p$ is very important as different structure in 
speech signals (our main case
 study) is evident at different time scales.
Let $I=I(\gamma_{\bar m,\bar l})=I(\mathcal F X_{\gamma_{\bar m, \bar l}})$ 
be a function defined for each path 
$\gamma_{\bar m,\bar l} \in \mathcal C_p$. 
We define now  {\it a semi-local thresholding estimator} in the window Fourier
 frame as follows:
\begin{equation}
\tilde F=\frac{1}{N}\sum_{m=0}^{N-1} 
\sum_{l=0}^{N-1} d_{I,T}(\mathcal F X[m,l])\tilde g_{m,l}
\end{equation}
where $d_{I,T}(\mathcal F X[m,l])=\mathcal F X[m,l]$ if 
$I(\mathcal F X_{\gamma_{\bar m,\bar l}})\geq T$ 
for some $\gamma_{\bar m,\bar l}$ 
containing $(m,l)$,
and $d_{I,T}(\mathcal F X[m,l])=0$ if $I(\mathcal F X_{\gamma_{\bar m,\bar l}})< T$ for all 
$\gamma_{\bar m, \bar l}$ containing $(m,l)$.
\\
\\
Note that this threshold estimator is build to mirror the diagonal 
estimators in (1), but that the `semilocal' 
quality of $\tilde F$ is evident from the fact 
that all coefficients 
in several $\mathcal F X_{\gamma}$ are used to decide the action of the 
thresholding on each coefficient. 
This procedure is similar to block thresholding estimators,
with the additional flexibility of choosing the index function $I$.
We propose in the next section a novel use of embedding 
techniques from 
non-linear dynamical systems theory  
to choose a specific form for $I$. We find in this way
a variance independent estimator 
that does not depend significantly on
the probability 
distribution of the random variable $W$ and such that 
we can adapt to the data in a flexible way.
\\
\\
\section{Delay-Coordinates Embedding Images of Time Series}

We first recall a fundamental 
result about reconstruction 
of the state space
realization of a dynamical system from its time series measurements.
Suppose $S$ is a dynamical system, with state space $\R^k$  and 
let $h:\R^k\rightarrow \R$ be a measurement, i.e., a continuous
function of the state variables. Define moreover a function $F$ of 
the state variables $X$ as
\begin{equation}
 F(X)=[h(X),h(S_{-\tau}(X)),...,h(S_{-(d-1)\tau}(X))]
\end{equation}
where by $S_{-j\tau}(X)$ we denote the state of the system with 
initial condition $X$ at $j\tau$ time units earlier.
\\
\\
We say that $A\subset \R^k$ is an invariant set with respect to 
$S$ if $X\in A$ implies $S_t(X)\in A$ for all $t$.
Then the following theorem is true (see [ASY], [SYC] and [KS]):
\\
\\
\\
{\bf Theorem:} {\it Let $A$ be an $m$-dimensional submanifold 
of $\R^k$
which is invariant under the dynamical system $S$. If $d>2m$, then 
for generic measuring functions $h$
and generic delays $\tau$, the function $F$ defined in (7) 
is one-to-one on $A$.} 
\\
\\
Keeping in mind that generally the most significant information about $g$ 
is the knowledge of the
attractive invariant subsets, we can say that delay maps allow to have a 
faithful description of
the underlining finite dimensional dynamics, if any. 
The previous theorem can be extended to invariant sets $A$ that are not 
topological manifolds; in that case more
sophisticated notions of dimension are used (see [SYC]).
\\
Generally the identification of the `best' $\tau$ and $d$ that allows for a
faithful representation of the invariant subset is considered very important in 
practical applications (as discussed in depth in [KS]), as it allows to make transparent the properties of 
the invariant set itself, more particularly we want to deduce from 
the data itself the dimension $m$ of the invariant set (if any) so that we can 
choose a $d$ that is large enough for the theorem to apply. Moreover 
the size of $\tau$ has to be large enough to resolve the image far from the 
diagonal, but small enough to avoid decorrelation of the delay coordinates
point.
\\
We apply the structure of the embedding in such a way that the identification of the most suitable $\tau$ and $d$
is not so 
crucial , even though
we will see that we do need to train such parameters on the available data, 
but in a much simpler and straightforward way. The technical reason for such 
robustness in the choice of parameters will be clarified later on,
 but essentially
we use time delay embeddings as {\it data mining tools rather than 
modelization tools as usually is the case.} 
\\
\\
To understand how such data mining is possible, we start by 
applying  the delay-coordinate procedure to the time series 
$W[n]$, $n=1,...,N$, for $W$
an uncorrelated random process; let the measuring function $h$ be the identity function and assume from now on that 
$\tau$ is an integer delay so that $F(W[n])=[W[n],W[n-\tau],...,W[n-(d-1)\tau]]$.
For any embedding dimension $d$, the state space will be filled
 according to a spherically symmetric probability distribution. Let now
$\bar Z=\{ F(Z[n]),\,\,n=1,...,N\} $ be the embedding image in $\R^d$ of a time series $Z$  for any given time delay $\tau$.
Then we have the following very simple, but fertile lemma that relates spherical distributions to 
their associated to principal directions
\\
\\
{\bf Lemma 1:} {\it Let $\sigma_1$, 
$\sigma_d$ be the variance of $\bar W$ along the first principal direction (of largest extent) and the last one (smallest) 
respectively. Then the expected value $E\{\frac{\sigma_1}{\sigma_d}\}$ converges to $1$ as $N$
goes to infinity.}
\\
{\it Proof:} Because $W$ is a white noise process, 
each coordinate
of $F(W[n])$ is a realization of a same random variable 
with some given 
probability density function $g$, therefore $\bar W$ 
is a realization of a multivariate random variable 
of dimension $d$ and symmetric probability distribution.
If the expected value of $\frac{\sigma_1}{\sigma_d}=Q>1$,
then a point at a distance from the origin of $\sigma_1$ has 
a greater probability to lie along the principal direction
associated to $\sigma_1$ contradicting the fact that the 
probability distribution of $\bar W$ was symmetric.
\\
\\
\\
{\bf Remark 1:} Even when $X$ is a pure white noise process, 
the windowed Fourier frame
will enforce a certain degree of smoothness along each path $\gamma$ since 
consecutive points in $\gamma$ are inner products of frame atoms 
with partially
overlapping
segments of $X$. 
So there will be some correlation in $\mathcal F X_{\gamma}$
even when $X$ is an uncorrelated time series, 
therefore {\it it is possible in general
that $I(\mathcal F X_{\gamma})>>1$ } even when $X$ is a 
white noise process. 
\\
\\
{\bf Remark 2:} Similarly, the length $p$ of $\gamma$ cannot 
be chosen very large 
in practice, while  $E(\frac{\sigma_1}{\sigma_d})$ converges to 
$1$ for {\it any} uncorrelated processes
only asymptotically for very long time series and again for small
length $p$ we may have $E(\frac{\sigma_1}{\sigma_d})>>1$. 
\\
\\
Even with the limitations explained in the previous two remarks, 
it is still meaningful to set 
$I(X_{\gamma})=I^{svd}(X_{\gamma})=\frac{\sigma_1}{\sigma_d}$, and
therefore we define {\it an embedding threshold estimator to be 
a semilocal estimator $\tilde F$ (as in (2)) with the choice 
of index $I=I^{svd}$, what we call an embedding index.}
The question is now to find a specific choice of $T \geq 1$,  given 
a choice of $(\mathcal D, C_p, d, \tau)$, that 
allows to discriminate a given data set (speech signals in this paper)
from white noise processes.
\\
\\
We need therefore to study the value distribution of $I^{svd}$ for our 
specific choice of $\mathcal C_p$ and $\mathcal D$, and 
assuming $X$ is either an uncorrelated random process or a signal belonging 
to our class of speech signals. 
\\
\\
In the next section we explore numerically this issue for the windowed 
Fourier frames and the collection of paths $C_p$ in (5).
\\
\\

\section{Embedding Index of Speech Signals and Random Processes}

For a given times series $X$ and choice of parameters $(p,\tau,d)$
we can compute the collection of embedding indexes
 $I^{svd}(\mathcal F X)=\{ I^{svd}(\mathcal F X_{\gamma}),\,\, \gamma \in C_p\}$, 
 Define now the {\it index cumulative function} as
\begin{equation}
Q_{X}(t)=\frac{\#\{\gamma \,\,\text{such that}\,I^{svd}(\mathcal F X_{\gamma})>t\}}{\#\{\gamma\}},
\end{equation}
i.e. for a given $t$, $Q_{X}(t)$ is the fraction of paths that 
have index above $t$.
\\
\\
A simple property  of $Q_{X}$ will be crucial in the following discussion:
\\
\\
{\bf Lemma 2:} {\it If $X$ is a white noise process and $X'=aX$ is another
random process that component by component is a rescaling of 
$X$ by a positive number $a$, then the expected function $Q_X$ and $Q_{X'}$ are equal.}
\\ 
{\it Proof:} Each set of embedding points generated by one specific 
path $\gamma$ is, coordinate by coordinate, a linear combination of some 
set of points in the  original time series. Therefore if $X'=aX$, 
$\bar {\mathcal F X'_{\gamma}}=a \bar {\mathcal F X_{\gamma}}$, but the quotient of singular values of a set of 
points is not 
affected by rescaling of all coordinates, therefore the distributions of $I^{svd}(\mathcal F X)$
and $I^{svd}(\mathcal F X')$ are equal, 
but $Q_{X'}$ and $Q_X$ are defined in terms of $I^{svd}$ so they are equal as well. 
\\
\\
\\
{\bf Remark 3:} We see the use of embedding index 
as a possible generalization of 
methods like the coherent structures extraction of 
[M] section 10.5 (more details can be found in [DMA]), 
where it is explored the notion of correlation 
of a signal $X$ of length $N$
with a basis $B$, defined as 
$$\mathcal C(X)=\frac{sup_{0\leq m<N}|X_B[m]|}{|X|}.$$
It turns out that in the limit $N\rightarrow \infty$ the correlation of 
any Gaussian white process converges to 
$$\mathcal C_N=\frac{\sqrt{2log_eN}}{\sqrt{N}}$$
{\it independently of the specific variance} and therefore estimation of a signal $X$
is performed by retaining a coefficient $X_B[m]$
if $\frac {|X_B[m]|}{|X|}>\mathcal C_N$. In this paper the 
embedding index
determines the  coherence of a coefficient with respect 
to a neighbourhood 
of the signal and it is independent of the variance 
of the noise process as well.
\\
\\
\\
{\bf Remark 4:} As we said in section II, the choice of 
$p$ in $C_p$ is very important 
in practice. 
The speech signals that we consider are 
sampled at a sampling frequency of  about 8100 pt/s, we 
choose supprt of the window $q=64$ and length of the paths $p=2^8$, since 
these values seem to assure  that each path will be significantly shorter than most 
stationary vocal 
emissions, a point to take into consideration when we gauge 
the relevance of our results.
\\
Given this lenght $p$ for $\gamma$, we have some significant 
restrictions
on the maximum embedding dimension $d$ and time delay $\tau$ that 
we can choose if we want to have for each path a sufficiently large 
number of points  in the embedding image to be 
statistically significant, which we can obtain if $p>>d\tau$.
\\
Because of these restrictions we choose 
$d=4$ and $\tau=4$ that give $d\tau=2^4<<p=2^8$, we generate in this way
240 points for each path. We heuristically tried to 
adjust the embedding parameters $d$ and $\tau$ and 
the lenght $p$ of the paths so that the qualitative behaviour of 
speech signals and white noise processes was as distinct as possible,
see the discussion in section IV for a possible way to 
make the choice of parameters automatic.
\\
\\
\\
We now expand some uncorrelated zero mean random processes 
of length
$N=2^{11}$ on the windowed Fourier 
frame with the set values $q=64$, $p=2^8$, $d=4$ and $\tau=8$. 
And we compute the embedding index  $Q_X$.
\\
The specific random processes we use here  
are time series with each point a realization of a 
random variables with: 
\\
1) Gaussian probability density function. 
\\
2) Uniform probability density function. 
\\
3) Tukey probability density function, that is, a sum of two 
normal distributions with uneven weight (used in [ELPT] as well), 
each point of the time series
is a realization of the random variable $W=RN_1+(1-R)4N_2/\sqrt{r+16(1-r)}$, 
where $N_1$ and $N_2$ are Gaussian random 
variables, and $R$ is a Bernoulli random variable with $P(R=1)=0.9$ and 
$r=P(R=1)$.
\\
4)discrete uniform pdf with  values in $\{-Q,Q\}$ for some 
positive $Q$.
\\
All probability density functions are set to have mean zero. 
and variance 1, since by Lemma 2 we know $Q_*$ will not be affected
by changes of the variance. One of the pdf has heavy tail (Tukey pdf)
and one of them is discrete (discrete uniform pdf). The 
kurtosis is respectively from pdf in 1) to pdf in 4):
$3$, about $1.8$, about $13$,and about $1.2$
\\
\\
In Figure 1a we plot $Q_{X}(t)$  for the white noise processes generated
with pdfs in 1)-4), averaged over 10 repetitions for each random
distribution. 
\\
\\
{\bf Remark 5:} To speed up the computation, we sampled the indexes 
$(\bar m,\bar l)$ of the paths in (5), more particularly we  
selected a sampling length of $S_{\bar m}=1$ for the frequency 
index $\bar m$ and  a sampling length of $S_{\bar l}=p$ for the time
index. 
\\
\\
Note that the qualitative behaviour of $Q_{X}$ 
is very similar for all chosen 
distributions, in
particular they all exhibit a very fast decay for larger values of $t$. 
The maximum $L_2$ distance between any two $Q_X$ in the interval $[0,40]$ 
is $\approx 0.54$ (or some $6\%$ of the average $L_2$ norm of the $Q_X$) ,
we found  that even for distribution with kurtosis up to $50$
the maximum distance was less that $0.8$ (about $8.5\%$ of the average $L_2$
norm of $Q_X$), irrispective of the specific pdf, moreover most 
of the error is concentrated in regions of high intensity of the derivative 
and it does not affect much the behaviour of the right tail of the curves $Q_X$. 
\\
\\
Therefore it seems that, for our choice of $\mathcal D$ and $C_p$,
reasonably heavy tail distributions will not exibit a 
significantly different 
behaviour in $Q_X$ with respect to the Gaussian distribution, supporting 
our claim that $Q_X$ is robust
with respect to the choice of white noise distribution. 
\\
\\
For each probability density function,  the shape of 
$Q_X$ is affected by
the correlation introduced by the length of $q$ 
(the window support of the windowed 
Fourier Frame): if $\tau<q$ some coordinates
in each embedding point will be correlated and this will 
cause the decay of $Q_X$ to be slower when $\tau$ 
is smaller.
\\
\\
\begin{figure}
\includegraphics[angle= 0,width=0.3\textwidth]{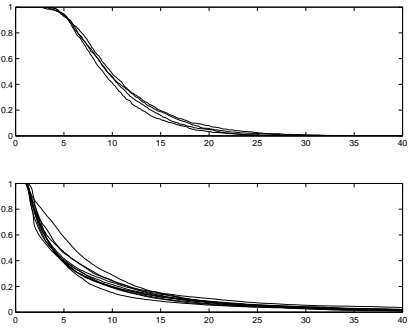}
\caption{From top to bottom, this figure shows $Q_{*}$, as defined in equation (7) for: a) uncorrelated 
random processes 1)to 4); b) ten randomly selected
segment of speech signal from the TIMIT database.}
\end{figure}
When $Q_X$ is computed (with the same choice of parameters) 
for a collection of $10$ randomly selected 
segments of 
speech signals of length $2^{11}$, the rate of decay of the functions 
$Q_{X}$ is  
significantly different, and the tail of the functions is 
still considerably 
thick
by the time the rate of decay of $Q_X$ for most random processes 
is almost zero (see Figure 1b).
\\
\\
Since we want to have a significantly 
larger fraction of paths retained
for speech signals rather than noise,  
we can select the 
threshold $T$ in the following way:
\\
\\
{\bf (A) Determination of Threshold} {\it Given a choice of parameters  
$(\mathcal D, \mathcal C_p, p, \tau,d)$, a 
 collection of training speech  time series $\{S_j\}$, 
and a selection of 
white noise processes $\{W_i\}$, choose $T_0$ to be the smallest $t$
so that the mean of 
$Q_{S_j}(T_{0})$ is one order of magnitute (10 times) larger than   
the mean of $Q_{W_i}(T_{0})$.} 
\\
\\
This heuristic rule gives, for the parameters in this section, 
$T_0\approx 28.2$. (A) gives us as experimental way 
to determine a threshold 
$T=T_{0}$ for the index 
$I^{svd}$ that 
removes most of 
the time frequency structure of some {\it predetermined} 
noise distributions, while it preserves a larger
fraction of the time frequency structure of speech signals. 
Since moreover `reasonable' distributions exibited a $Q_X$
similar to the one of Gaussian distributions, we can in practice 
train the threshold only on Gaussian noise and be assured that 
it will be a meaningful value for a larger class of distributions.
\\
\\
Note that even very low energy paths could have in principle 
high 
embedding index, still, the energy concentration in paths that 
have very 
high index tends 
to be large for speech signals, to see that, for a given signal 
$X$, let 
\begin{equation}
E_{X}(t)=\frac{\sum \{|\mathcal F X_{\gamma}|_2 \,\,\text{such that}\,\,I^{svd}(\mathcal F X_{\gamma})>t\}}{\sum|\mathcal F X_{\gamma}|_2},
\end{equation}
be the fraction of the total energy contained in paths with index above $x$.
We can see in Figure 2 that the amount of energy contained in 
paths with high index value
is significantly larger for speech signals  than for noise 
distributions.
\\
\\
More particularly, the fraction of the total energy of the paths 
carried 
by paths with 
$I^{svd}>T_{0}$ is on average $0.005$  for the  noise distributions
and $0.15$  for the speech signals, or an increase by a factor of $30$.
\\
\\
\begin{figure}
\includegraphics[angle= 0,width=0.3\textwidth]{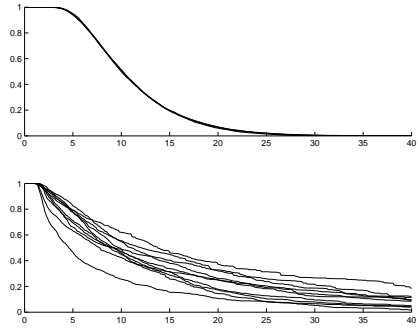}
\caption{ From top to bottom, this figure shows $E_{*}$, as defined in equation (8) for: a) the uncorrelated 
random processes in Figure 1a; b) 
the segments  of speech signals in Figure 1b.}
\end{figure}
It seems therefore that  $I^{svd}$, with our specific choice of 
parameters, 
is quite
 effective in separating  a subset of paths that are likely to be generated by  
speech signals, note moreover that 
similar results can be obtained with 
local changes of  $p$, $\tau$ and $d$, which suggests an intrinsic robustness of 
the separation with respect of the parameters.
\\
This separation ability could be due, in principle, only to the 
very nice properties of speech signals.
 Note that if, for some $\mathcal F X_{\gamma}$,  $I^{svd}=\infty$,
 then the state 
realization of the time
 series $\mathcal F X_{\gamma}$ is embedded in a subspace of 
$R^d$ and therefore 
each point of   
 $\mathcal F X_{\gamma}$ must be described as a linear function of the delay 
coordinates. This condition 
is very restrictive on the dynamics of  $\mathcal F X_{\gamma}$, but vocal 
emissions are locally
 periodic signals, and so they do fall, at least locally, 
into the class of linearly
 predictable discrete models, i.e., processes
for which $X_k=r(X_{k-1},...,X_{k-d})$ for some linear 
function $r$ and for some integer $d$. 
\\
The complexity of these linear models increases with increasing 
values of the embedding dimension $d$. 
But this is not fully satisfactory as we would like 
to be able to use the embedding index $I^{svd}$ to  
denoise more complex dynamics that cannot be described 
by simple linear 
predictive models.
Moreover for small $\tau$ we are measuring in many cases smoothness
of the path and local correlation with the embedding index, yet,
if we try to choose $\tau$ as large as possible with still a clear 
separation of the training sets, we can see differences that are not 
accounted for by local correlation, indeed the embedding image 
is squeezed along the diagonal for paths with high local smoothness,
but in principle for complex dynamics the principal direction could be 
oriented in any direction and therefore the embedding index 
is much more than simply a measure of local smoothness.
\\
\\
There is a large literature on possible ways to distinguish 
complex dynamical systems  from random behaviour 
(see for example the articles collected in [Me]), as we underlined in the previous section,
much of this work stresses the 
identification of the proper embedding parameters $\tau$ and $d$;
the contribution of this paper
to this ongoing discussion is  the use of embedding 
techniques in the context of computational harmonic analysis.
This context frees us from the 
need to use embedding techniques to find an effective 
modelization of the signals, such `blind' use of 
the embedding theorem is, we believe, fertile from a practical point of view, as 
well as a theoretical one.
\\
\\
Note in any case that if the dimension of the invariant set $A$ is $d_A=0$, then for any white noise process
$W$, $X+W$ has spherically symmetric embedding image and $\frac{\sigma_1}{\sigma_d}\approx 1$
for any embedding dimension $d$ as in the case of pure white noise. This means that 
an estimator based on $I^{svd}$ is not able to estimate noisy constant time series on a 
given path $\gamma$. This
restriction can be eased by allowing information on the distance of the center
of the embedding image to be included in the definition of 
the embedding threshold estimator. 
In this paper for simplicity we assumed $d_A>0$ for all paths in $C_p$. 
That seems to be sufficient in analyzing speech signals.

\section{Attenuated Embedding Estimators}

In this section we develop an algorithm 
based on these ideas. The notion of semilocal estimator is slightly expanded
to improve the actual performance of the estimator itself.
To this extent, define  tubular 
neighborhoods for each atom in the windowed Fourier
frame, i.e.:
\begin{equation}
\mathcal O(g_ {m, l})=\{g_{m',l'} \,\, \text{s.t.}\,\, |l'-l|\leq 1,  |m'-m|\leq 1\},
\end{equation}
Such neighborhoods  are used in the algorithm as a way to
make
a decision on the value of the coefficients in a {\it two dimensional} 
neighborhood of
$\mathcal F X_{\gamma}$ based on the 
the analysis of  the 
{\it one dimensional} time series $\mathcal F X_{\gamma}$   itself. 
\\
\\
{\bf (C1)} {\it Set $\tilde F=0$.}
\\
\\
{\bf (C2)} {\it Given  $X$, choose $q>0$ and 
expand $X$ in a windowed Fourier frame with window size $q$.}
\\
\\
{\bf (C3)} {\it Choose  sampling intervals $S_{\bar l}$ for time coordinate and  $S_{\bar m}$ for the frequency coordinate.
Choose the  path length $p$. 
Build a collection of paths $\mathcal C_p$ as in (5).}
\\
\\
{\bf (C4)} {\it Choose embedding dimension $d$ and delay $\tau$ along the 
path. Compute the index $I^{svd}(\mathcal F X_{\gamma_{\bar m,\bar l}})$
for each $\mathcal F X_{\gamma_{\bar m,\bar l}}\in \mathcal C_p$. Use (A) 
to find the threshold level $T$.}
\\
\\
{\bf (C5)} {\it Choose attenuation 
coefficient $\alpha$. Set $\mathcal F Y[m,l]=\alpha \mathcal F X[m,l]$ 
if $I^{svd}(\mathcal F X_{\gamma})\geq T$ 
for some $\gamma$ containing $g_{m',l'}$, 
$g_{m',l'}\in \mathcal O(g_{m,l})$,
otherwise set $\mathcal F Y[m,l]=0$ if $I^{svd}(\mathcal F X_{\gamma})< T$ 
for all 
$\gamma$ containing $g_{m',l'}$, $g_{m',l'}\in \mathcal O(g_{m,l})$.}
\\
\\
{\bf (C6)} {\it Let $Y$ be the inversion of 
 $\mathcal F Y$. Set $\tilde F=\tilde F+Y$ and $X=X-Y$.}
\\
\\
{\bf (C7)} {\it  Choose a paramenter $\epsilon>0$, if $|Y|>\epsilon$
go to step (C2).}
\\
\\
Note that the details of the implementation (C1)-(C7)
are in line with the general strategy of matching pursuit. 
The window length $q$ in step (C2) could change from one iteration
to the next to `extract' possible structure belonging to the
underlining signal at several different scales. In the experiments 
performed in the 
following section we alternate between two window sizes
$q_1$ and $q_2$.
\\
The attenuation introduced in (C5) has 
some additional ad hoc parameters in the definition of the 
neighborhoods in (10) and in the choice of the attenuation 
parameter $\alpha$. 
By the double process of increasing the number 
of nonzero 
coefficients chosen at each step
and decreasing their contribution we are allowing more information 
to be taken at each 
iteration of the projection pursuit algorithm, but in a slow 
learning framework that in principle (and in practice as we found out)
 should increase the sharpness of the distinct features 
of the estimate, on the general issue of attenuated learning processes
see the discussion in [HTF] chapter 10. {\it Note that the attenuation coefficient 
leads to improved results only when it is part of a recursive 
algorithm, otherwise it gives only a rescaled version of the estimate.}   
\\
\\ 
One drawback of the algorithm we described  
is the need to choose 
several parameters: 
we choose a dictionary of analysis $\mathcal D$, a collection of discrete 
paths $C_p$, the embedding parameters $\tau$ (time delay) and $d$ 
(embedding dimension), and the
learning parameters $T$ (threshold level), $\alpha$ (attenuation coefficient) and 
$\epsilon$.
Again we stress that all such choices are context dependent, and 
are the price to pay to have an estimator that is relatively 
intensity independent and
applicable to wide classes of noise distributions.
\\
The choice of $\mathcal D$ is dependent on the type of signals
we analyze and we do not see a serious need to make such choice automatic.
\\
Since we analyze speech signals, we 
choose the dictionary to be the set of atoms of the windowed Fourier frames; the algorithm is not 
very sensitive to the choice of the length $q$ of the window in the Fourier frame, while 
the use of several windows is found to be always beneficial. 
\\
The choice of  $C_p$ is also dependent on the type of signals analyzed, speech signals  have 
specific frequencies that change in time, so a set of paths parallel to the time 
axis was natural in this case. Let us explore now 
the relation of  parameters associated with $C_p$, 
embedding 
parameters $\tau$ and $d$ and threshold  $T$.
Recall that 
for the collection $C_p$ 
we have as parameters the time and frequency sampling 
rates $\bar l$ and $\bar m$ and the length $p$ of the paths. 
The frequency sampling 
rates $\bar l$ and $\bar m$ are necessary only to speed up the algorithm,
ideally we would like a dense sampling. Same considerations apply to 
the `thickening' of the paths in (10), we basically try to speed up the 
algorithm by collecting more data at each iteration.  
\\
So the only essential parameters are the path length $p$, the embedding 
parameters and the threshold $T$
\\
Essentially we want to set these parameters so that 
the number of paths that have index $I^{svd}>T$ is sizeable for 
a training set of speech signals and marginal for the white noise 
time series of interest. 
\\
Our experience is that 
such choice is possible and robust, we gave a simple rule to 
find the 
threshold $T$ in step (A) in the previous section given a choice of 
$(p,\tau,d)$. 
\\
A learning algorithm could be built to find $T$, 
the paths' length $p$, 
and the embedding parameters, namely let 
$\bar Q_S(x)$ be the mean of the functions
$Q_{S_i}(x)$ for a training set of speech signals $S_i$
and $\bar Q_W(x)$ be the mean of the functions
$Q_{W_i}(x)$ for a set of white noise time sieries $W_i$
\\
We can first find $d$, $\tau$ and $p$ such that 
the distance of the functions $\bar Q_W(x)$ and $\bar Q_S(x)$ 
is maximum in the $L^2$ norm. After finding these parameters, 
we can find a value of $T$ such that $T$ is the smallest 
positive number with $\bar Q_S(T)$ one order of magnitude
larger than $\bar Q_W(T)$, as we did in (A) in the previous 
section, to make our algorithm  automatically applicable to 
data sets of interest different from speech signals it will 
be necessary to formalize this optimization procedure.
\\
\\
Finally the choice of $\alpha$ and $\epsilon$ is completely practical 
in nature,
ideally we want $\alpha$ and $\epsilon$ as close to zero as possible, 
but, to avoid making
the algorithm unreasonably slow, we must set values that are found to 
give good quality reconstructions on some training set of speech signals
while they require a number of iterations of the algorithm that is compatible
with the computing and time requirements of the specific problem. 
For longer time series, as the ones in the next section, we 
segment the data in several shorter pieces, and we iterate the 
algorithm a fixed number of times $k$ rather than using  
$\epsilon$ in (C7) to decide the number of iterations.
\\
\\
{\bf Note:}{\it The algorithm described in this paper 
is being patented, with provisional patent application number 
60/562,534 filed on April 16, 2004.}

\section{Denoising}

In this section we explore  the quality of the attenuated
embedding threshold as implemented in the windowed Fourier 
frame and with our class of paths $\mathcal C_p$. 
We apply the algorithm to 10 speech signals from the TIMIT database 
contaminated by different types of white noise with several
intensity levels. 
We show that the attenuated embedding threshold estimator performs 
well for all white
 noise contaminations we consider.
\\
The delay along the paths  is chosen as $\tau=4$, the length 
of the paths is $p=2^8$ and the window length of the windowed Fourier transform 
alternates between $q=100$ and $q=25$ (to detect both features
with good time localization and those with good frequency localization), 
the  embedding dimension $d=4$.
For these parameters and for the set of speech signals that we
used as training, we have $T\approx 26.8$ when $q=100$ and $T\approx 27.4$ when 
$q=25$ using the procedure (A) of section III.
\\ 
The sampling interval of the paths in the frequency direction is 
$S_{\bar m}=3$ and along the time direction is $S_{\bar l}=p/2$
We select $\alpha=0.1$, as small values of $\alpha$ seem to 
work best (see discussion in the previous section).
The algorithm is applied to short consecutive speech segments to
reduce the computational cost of computing the windowed Fourier transform
on very long time series, therefore,
to keep the running time uniformly constant for all such segments,
we decided to iterate the algorithm (C1)-(C6) a fixed number of
 times (say 6 times)  instead of 
choosing a parameter $\epsilon$ in (C7).
\\
As we already said, the window size $q$ in (C2) alternates between $q=100$ and $q=25$.
It is moreover important to note that the attenuated embedding threshold 
is able to extract only a small fraction of the 
total energy of the signal $f$, exactly because of the attenuation process, 
therefore the Signal-to-Noise Ratio ($SNR$) computations are done on 
scaled measurements $X$, estimates $\tilde F$, and signals $F$ set to be
all of norm 1. We call such estimations {\it scaled} $SNR$, 
and we explicitely write, for a given signal $F$ and estimation $Z$,
$$SNR_s(Z)=10log_{10}\frac{1}{E(|F/|F|-Z/|Z|)}$$
We then compute $SNR_s(X)$ and $SNR_s(\tilde F)$ by approximating
the expected values $E(|F/|F|-X/|X|)$ and $E(|F/|F|-\tilde F/|\tilde F|)$
with an average over several realizations for each white noise contamination.
\\
\\
In Figure 3 we show the  gains of 
the scales SNR of the 
reconstructions  
(with the attenuated embedding threshold estimator)
plotted against the corresponding scaled SNR of the  measurements.
Each curve correspond to one of $10$ speech signals of approximately one second 
used to test the algorithm.
From top left in clockwise direction we have measuremets contaminated by 
random processes with pdfs 1) to 4) as defined in section III  and with 
several choices of variance.
\begin{figure}
\includegraphics[angle= 0, width=0.4\textwidth]{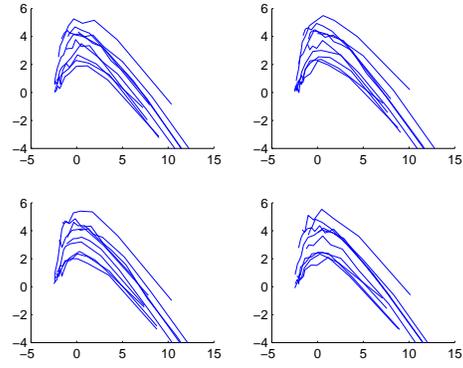}
\caption{Scaled SNR gain in decibel of the 
attenuated embedding estimates plotted against the scaled SNR
of the corresponding measurements. From top left in clockwise 
order we consider the case of: a)Gaussian white noise; b)
uniform noise; c)Tukey white noise; d)discrete bimodal distribution .}
\end{figure}
Note that the overall shape of the scaled SNR gain is similar for
all distributions (notwithstanding that the discrete 
plots do not have exactly the same domain). The maximum gain 
seems to happen for measurements with scaled SNR around $1$ decibel.
Note that the right tail of the
SNR gains takes often negative values; this is due to the attenuation 
effect of the estimator that is pronunced for the high intensity
speech features, but it is not necessarily indicative of worse
perceptual quality with respect to the measurements, some of the figures in the 
following will clarify this point. 
\\
\\
In the first case of Gaussian white noise, we compared our algorithm to the 
block thresholding algorithm described in [CS], we used the matlab code
implemented by [ABS], made available at $www.jstatsoft.org/v06/i06/codes/$ as a part
of their thourogh 
comparison of denoising methods.
As the block thresholding estimator is implemented in a
symmlet wavelet basis that is not well adapted to the structure of speech signals, a 
more compelling 
comparison would require the development of an embedding threshold estimator 
in a wavelet basis, we plan to do so in a future work. 
In Figure 10 we show the scaled SNR gain for all tested speech
signals using
the block threshold estimator (right plot) 
and attenuated embedding estimator (left plot).
In Figure 4 we show one original speech signal, 
Figure 5 shows the measurement in the presence of Gaussian
noise corresponding to the `peak' of the $SNR_s$ gain curve 
(measurement $SNR_s$ $\approx 1$), Figure 6 shows the 
corresponding
reconstruction with attenuated embedding threshold estimator.
Similarly Figure 7 shows another speech signal, while Figure 8
 shows the measurement  
with Tukey
noise corresponding to the `peak' of the Tukey noise 
$SNR_s$ gain curve 
(measurement $SNR_s$ $\approx 1$), 
Figure 9 shows the  
reconstruction. In both cases the perceptual quality 
is better than the noisy measurements, which is not 
necessarily the case for estimators in general.
\begin{figure}
\includegraphics[angle= 0, width=0.25\textwidth]{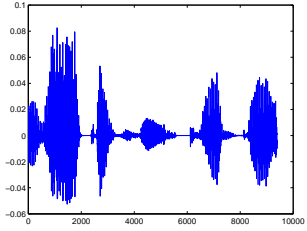}
\caption{Signal `SPEECH10' scaled to have norm $1$.}
\end{figure}
\begin{figure}
\includegraphics[angle= 0, width=0.25\textwidth]{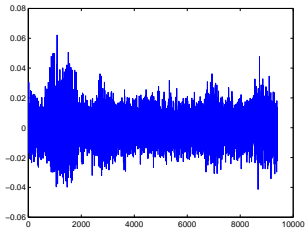}
\caption{Noisy scaled measurement of SPEECH10
 with Gaussian white noise
and scaled SNR of about $1$db.}
\end{figure}
\begin{figure}
\includegraphics[angle= 0, width=0.25\textwidth]{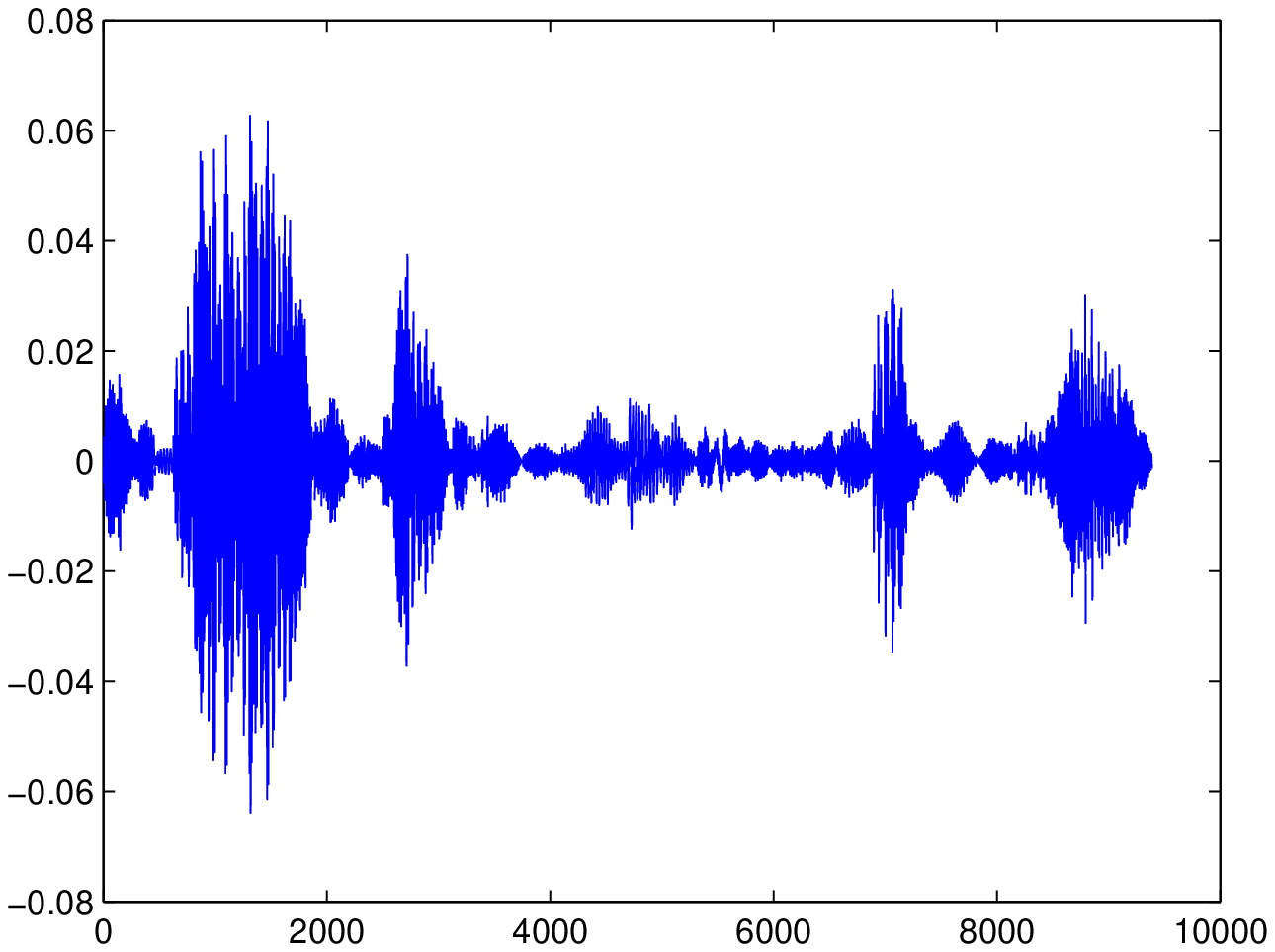}
\caption{Attenuated embedding estimate of SPEECH10 from the measurement 
in Figure 6, scaled to have norm 1.}
\end{figure}
\begin{figure}
\includegraphics[angle= 0, width=0.25\textwidth]{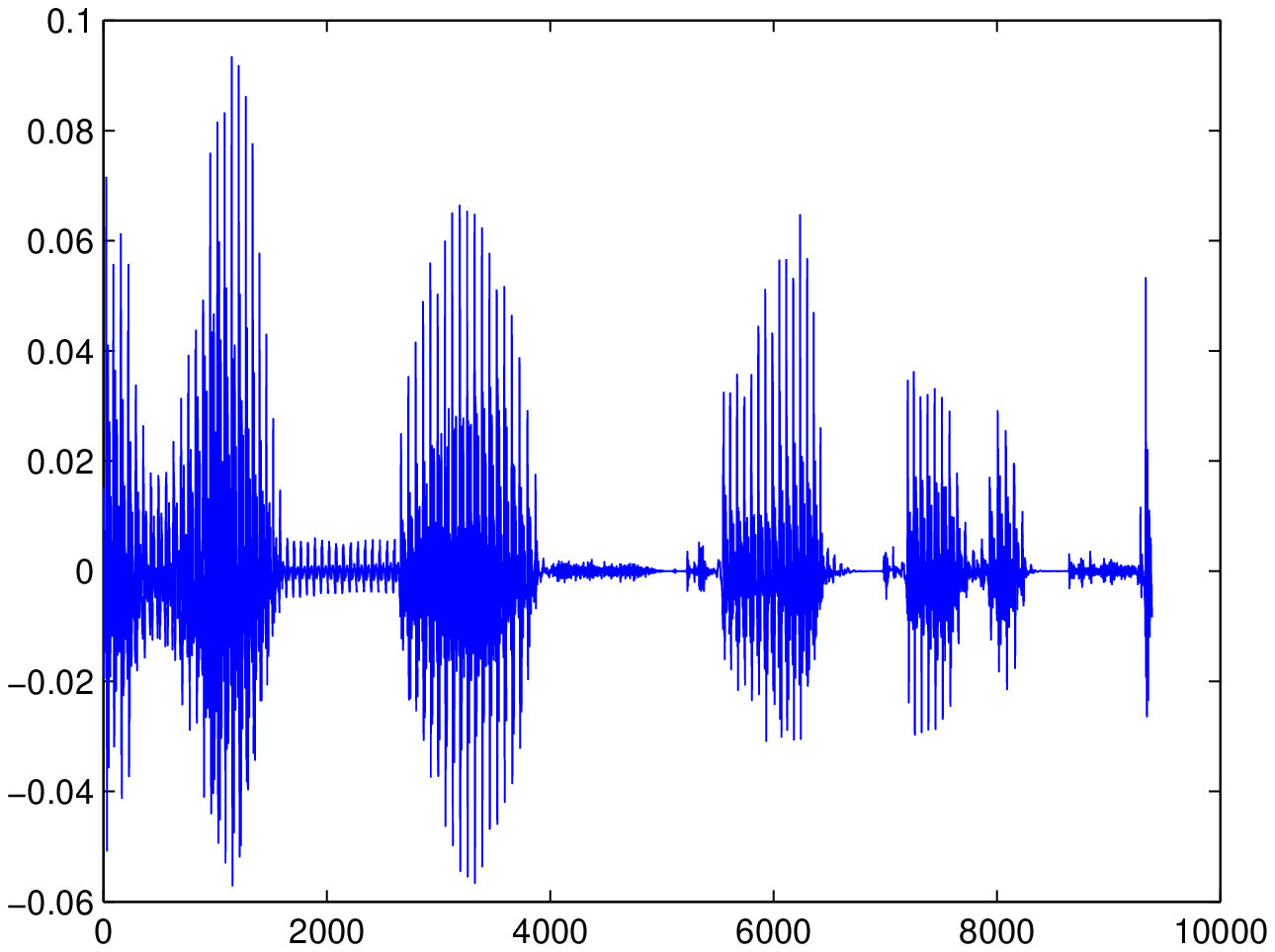}
\caption{Signal `SPEECH5' scaled to have norm $1$.}
\end{figure}
\begin{figure}
\includegraphics[angle= 0, width=0.25\textwidth]{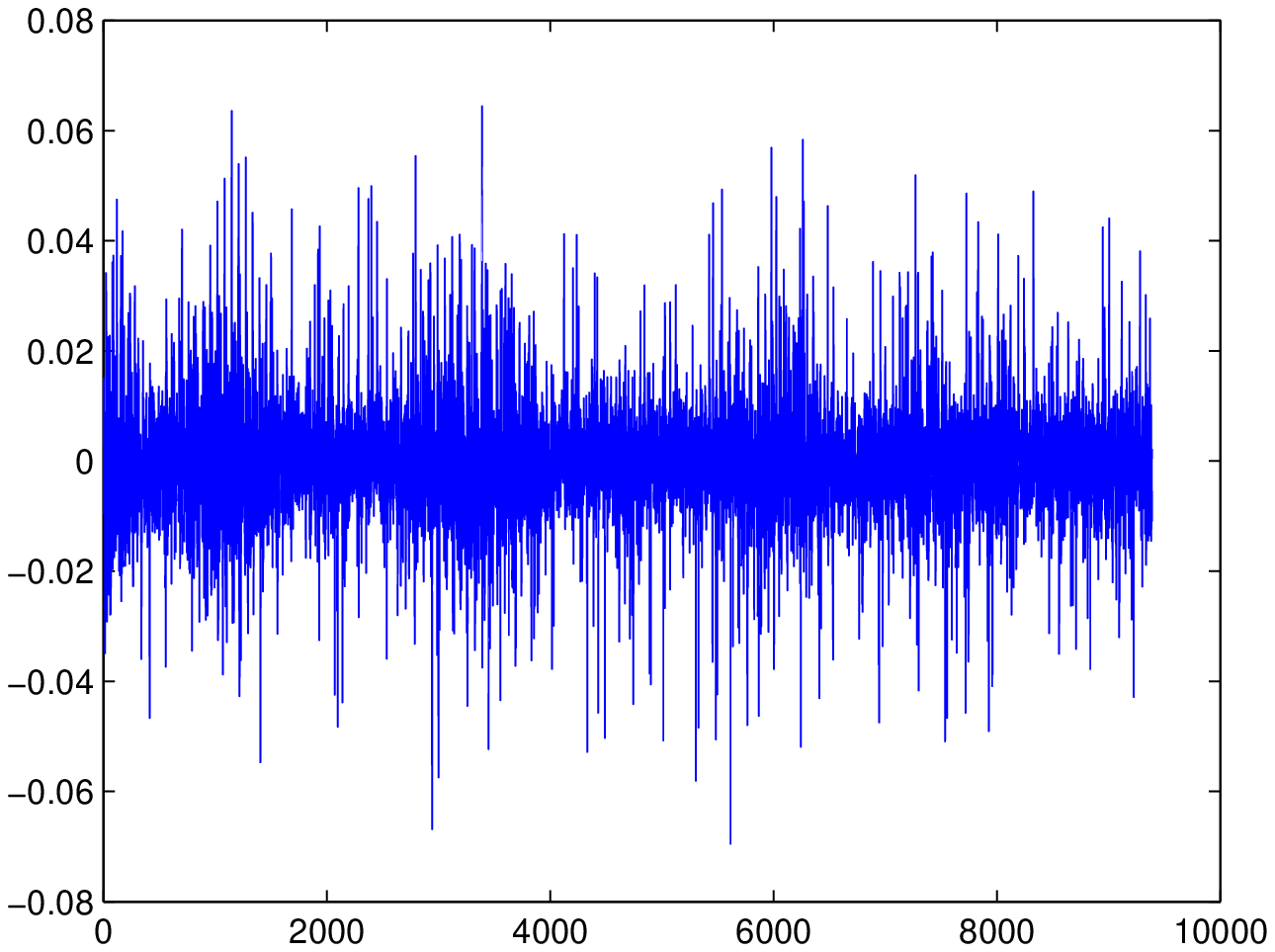}
\caption{Noisy measurement of SPEECH5
 with Tukey white noise 
and scaled SNR of about $1$db.}
\end{figure}
\begin{figure}
\includegraphics[angle= 0, width=0.25\textwidth]{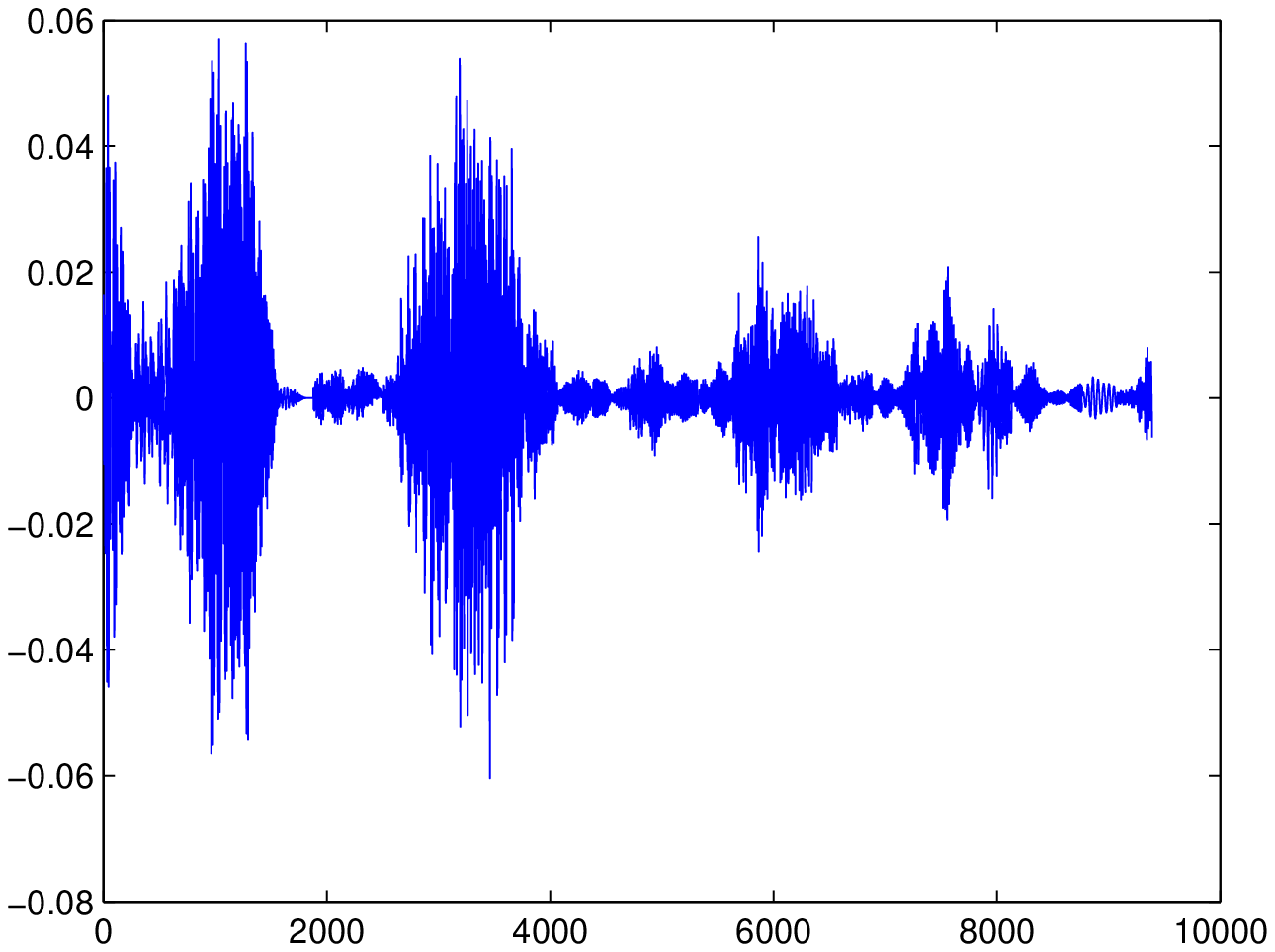}
\caption{Attenuated embedding estimate of SPEECH5
 from the measurement in Figure 9, scaled to have norm 1.}
\end{figure}
\begin{figure}
\includegraphics[angle= 0, width=0.25\textwidth]{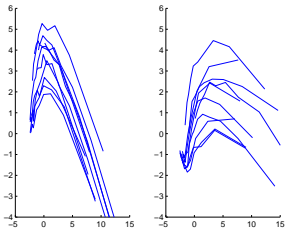}
\caption{$SNR_s$ gain for the estimates of 10 speech signals and Gaussian additive 
noise using: the block thresholding estimator of [CS](right),
the embedding threshold estimator(left).}
\end{figure}
\\
Note moreover that even though $T$ was found using only Gaussian 
white noise as the training distribution, none of the parameters of the 
algorithm were changed 
as we went from Gaussian white noise contaminations to  
more general 
white noise processes, and yet the $SNR_s$ gain was similar, it must be noted though
that the estimates for bimodal and uniform noise were not intelligible 
at the peak of the $SNR_s$ gain curve (just as the measurements were not). 
\\
\\
Since the performance of the embedding estimator is not well 
represented by the scaled SNR for low intensity noise (measurements appear
to be better than the estimates), in Figures 10 to 21
we show two more instances of speech signals contaminated by lower 
variance Tukey noise, Gaussian noise and discrete bimodal noise 
(uniform noise leads to
reconstructions very similar to the discrete bimodal distribution), for one case of low 
Gaussian white noise we show a block thresholding estimate, note how the low intensity details
are lost, this inability to preserve low intensity details worsens when 
higher variance noise is added, but then again, it must be tempered 
by the fact that a standard wavelet basis is not well adapted to the structure of
speech signals.
\begin{figure}
\includegraphics[angle= 0, width=0.25\textwidth]{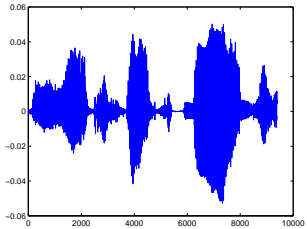}
\caption{Signal `SPEECH2' scaled to have norm $1$.}
\end{figure}
\begin{figure}
\includegraphics[angle= 0, width=0.25\textwidth]{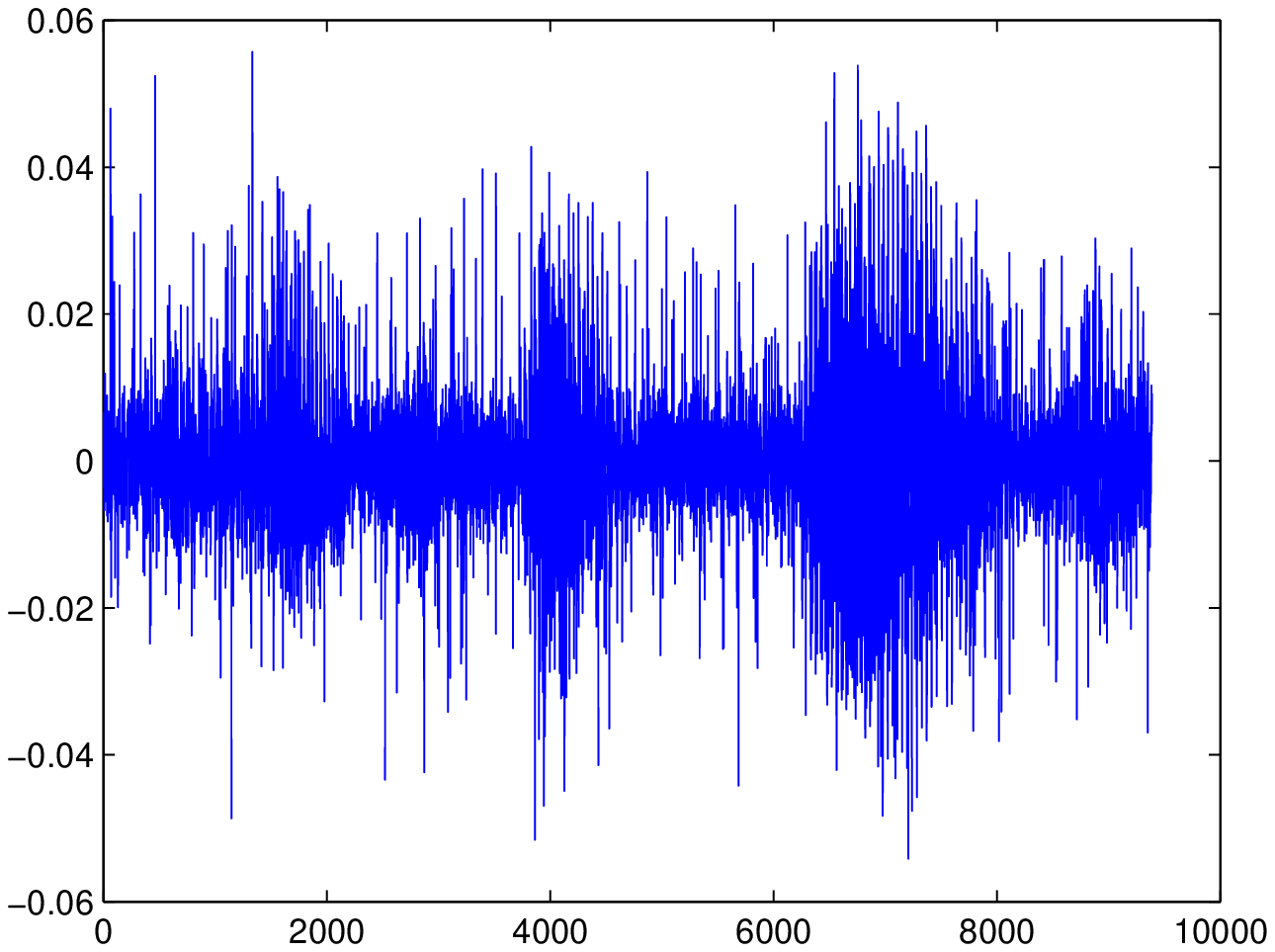}
\caption{Noisy measurement of SPEECH2 with Tukey white noise 
and scaled SNR of about $4.4$db.}
\end{figure}
\begin{figure}
\includegraphics[angle= 0, width=0.25\textwidth]{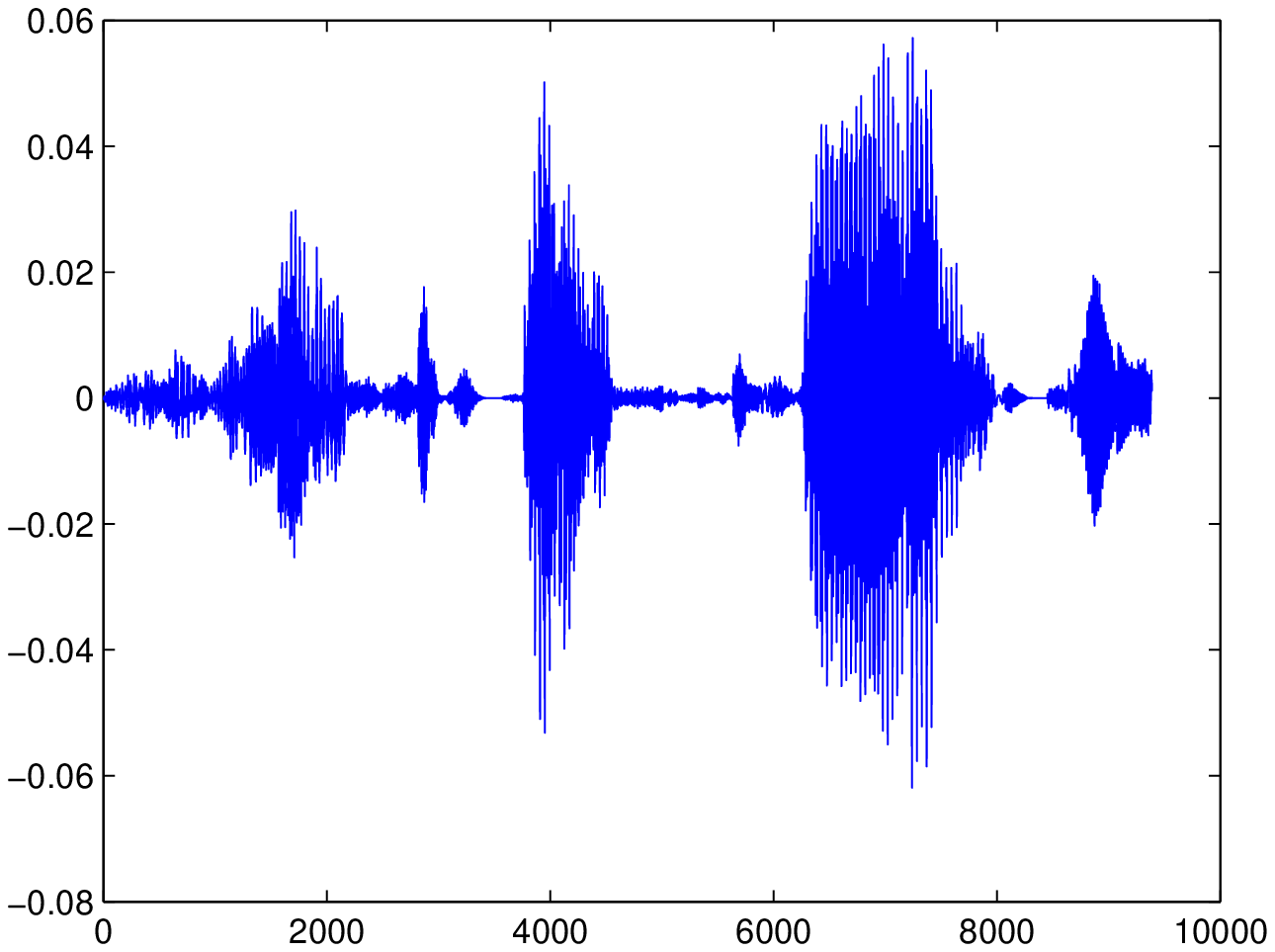}
\caption{Attenuated embedding estimate of SPEECH2
 from the measurement in Figure 12, scaled to have norm 1, $SNR_s$ is $\approx 8.1$db.}
\end{figure}
\begin{figure}
\includegraphics[angle= 0, width=0.25\textwidth]{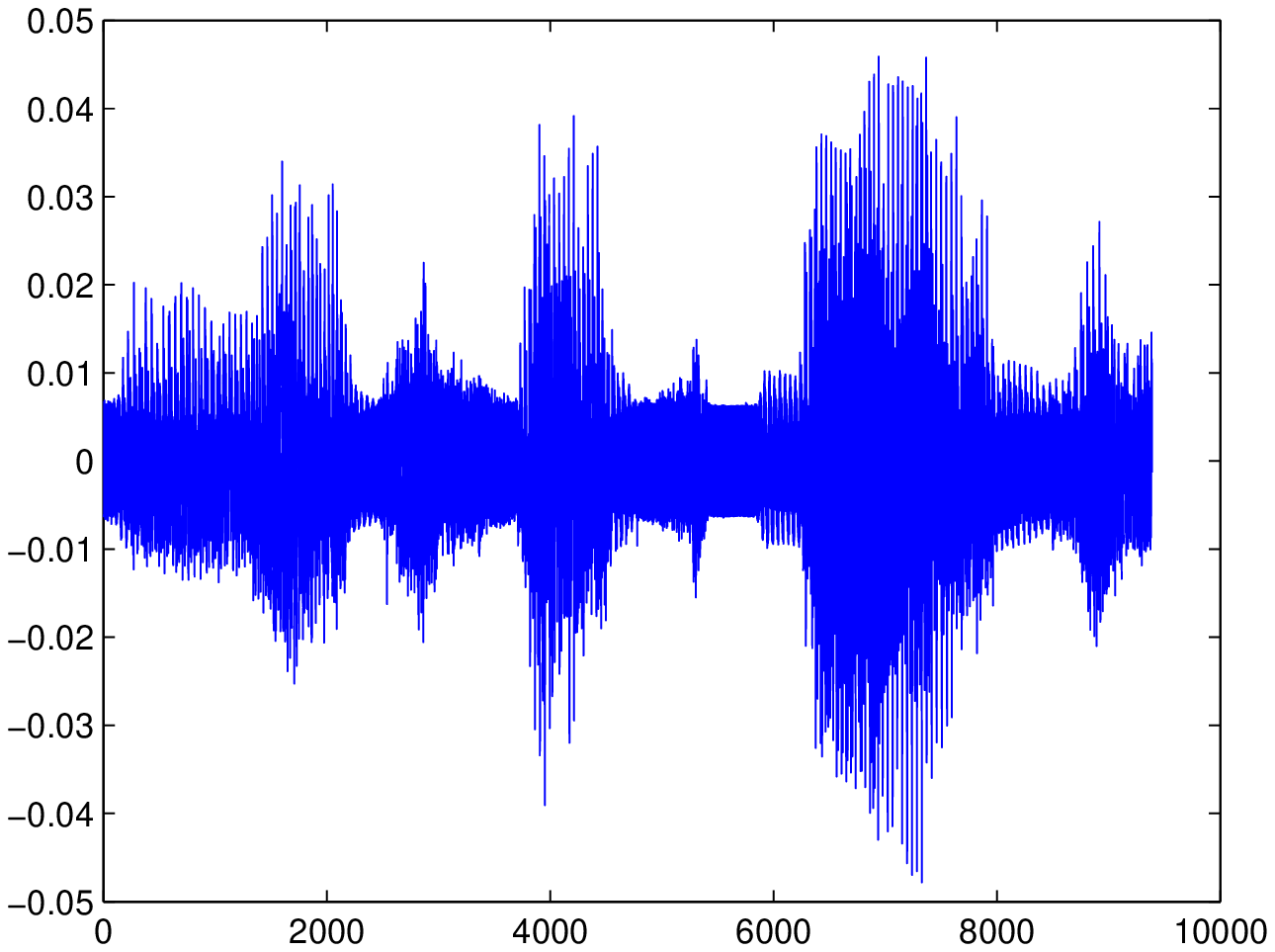}
\caption{Noisy measurement of SPEECH2
 with  bimodal white noise 
and scaled SNR of about $4.5$db.}
\end{figure}
\begin{figure}
\includegraphics[angle= 0, width=0.25\textwidth]{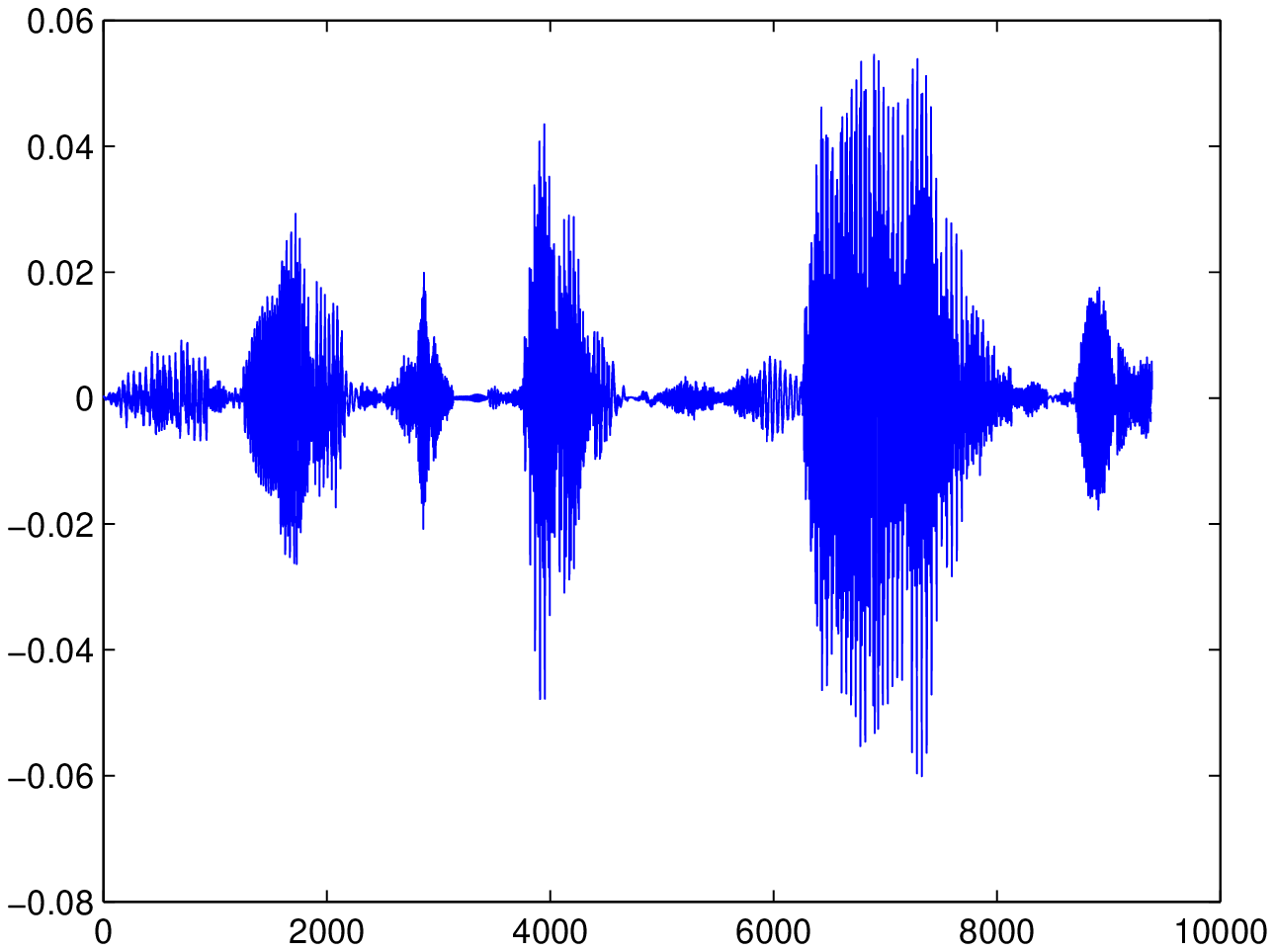}
\caption{Attenuated embedding estimate of SPEECH2 
 from the measurement in Figure 14, scaled to have norm 1, $SNR_s$ is $\approx 8.1$db.}
\end{figure}
\begin{figure}
\includegraphics[angle= 0, width=0.25\textwidth]{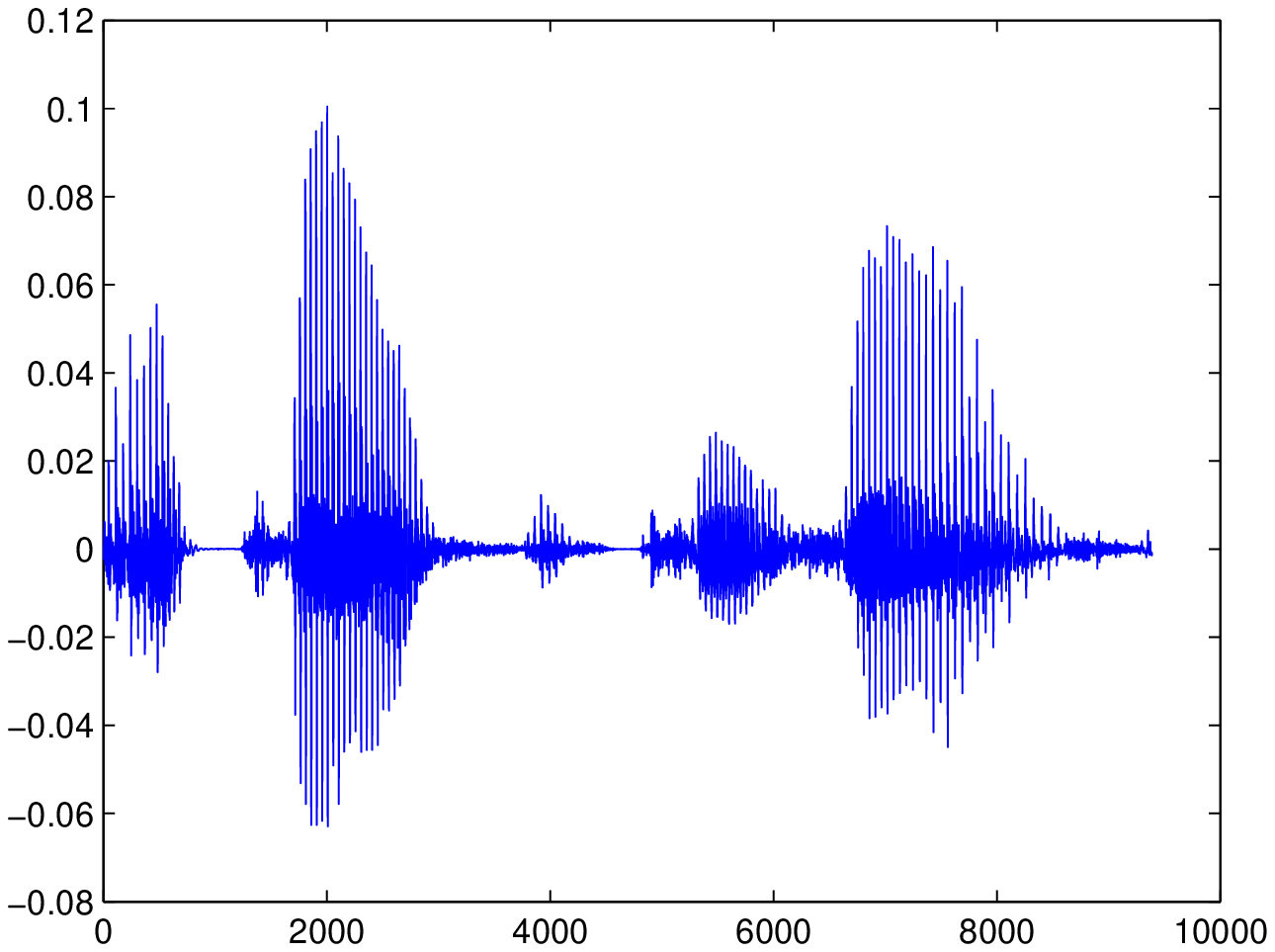}
\caption{Signal `SPEECH7' scaled to have norm $1$.}
\end{figure}
\begin{figure}
\includegraphics[angle= 0, width=0.25\textwidth]{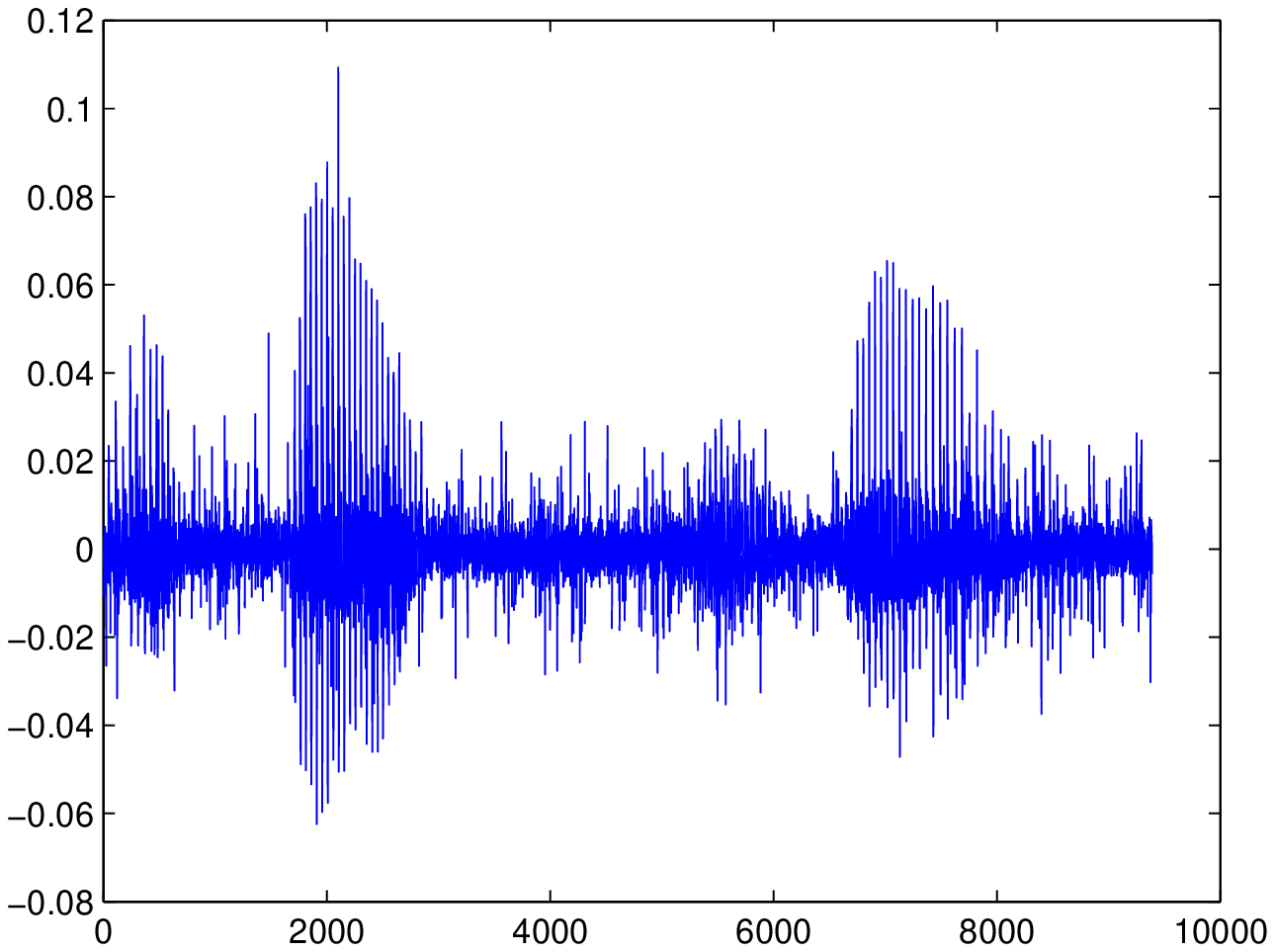}
\caption{Noisy measurement of SPEECH7
 with Tukey white noise 
and scaled SNR of about $7.3$db.}
\end{figure}
\clearpage
\begin{figure}
\includegraphics[angle= 0, width=0.25\textwidth]{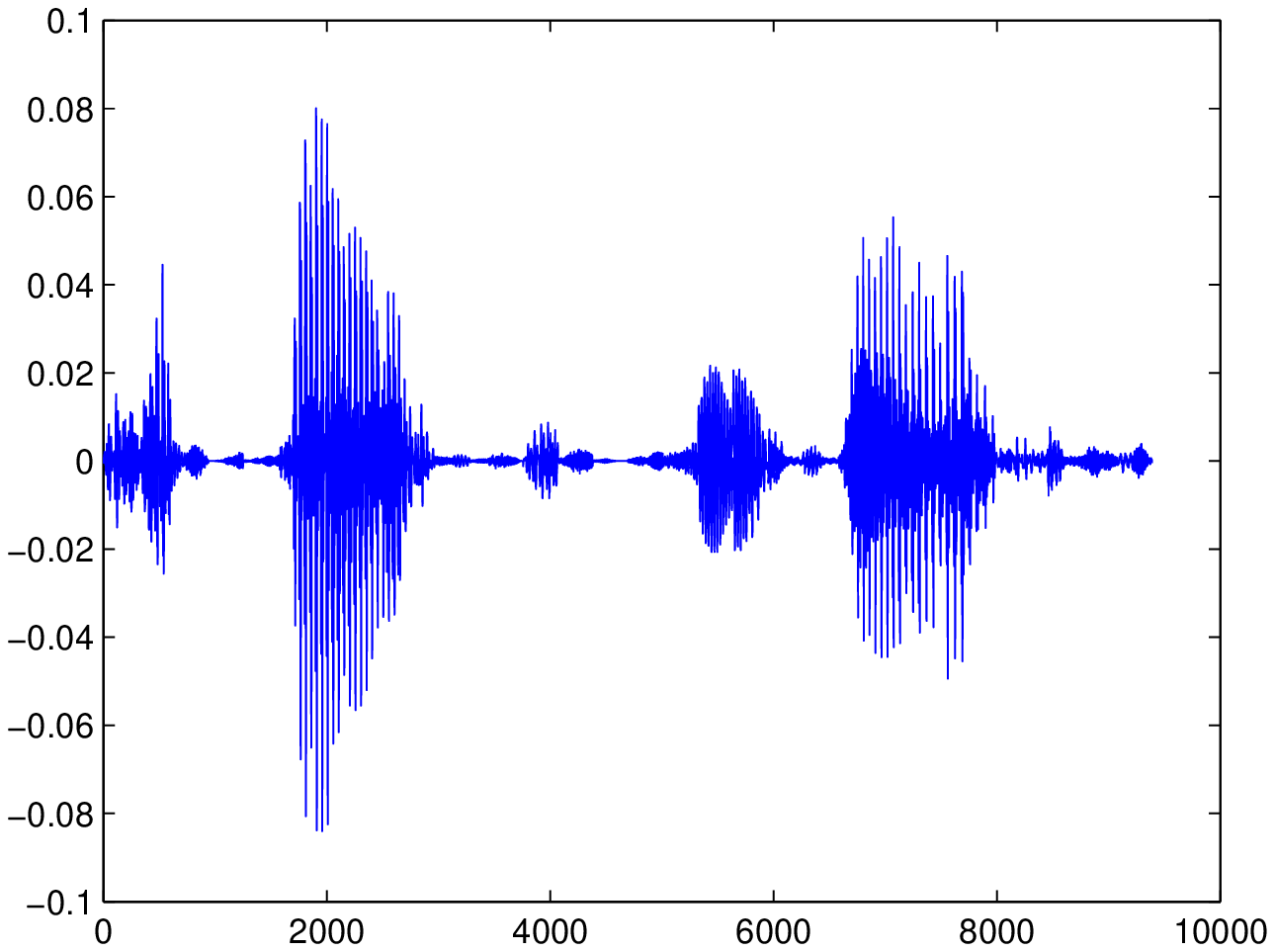}
\caption{Attenuated embedding estimate of SPEECH7
 from the measurement in Figure 17, scaled to have norm 1, $SNR_s$ is $\approx 6$.}
\end{figure}
\begin{figure}
\includegraphics[angle= 0, width=0.25\textwidth]{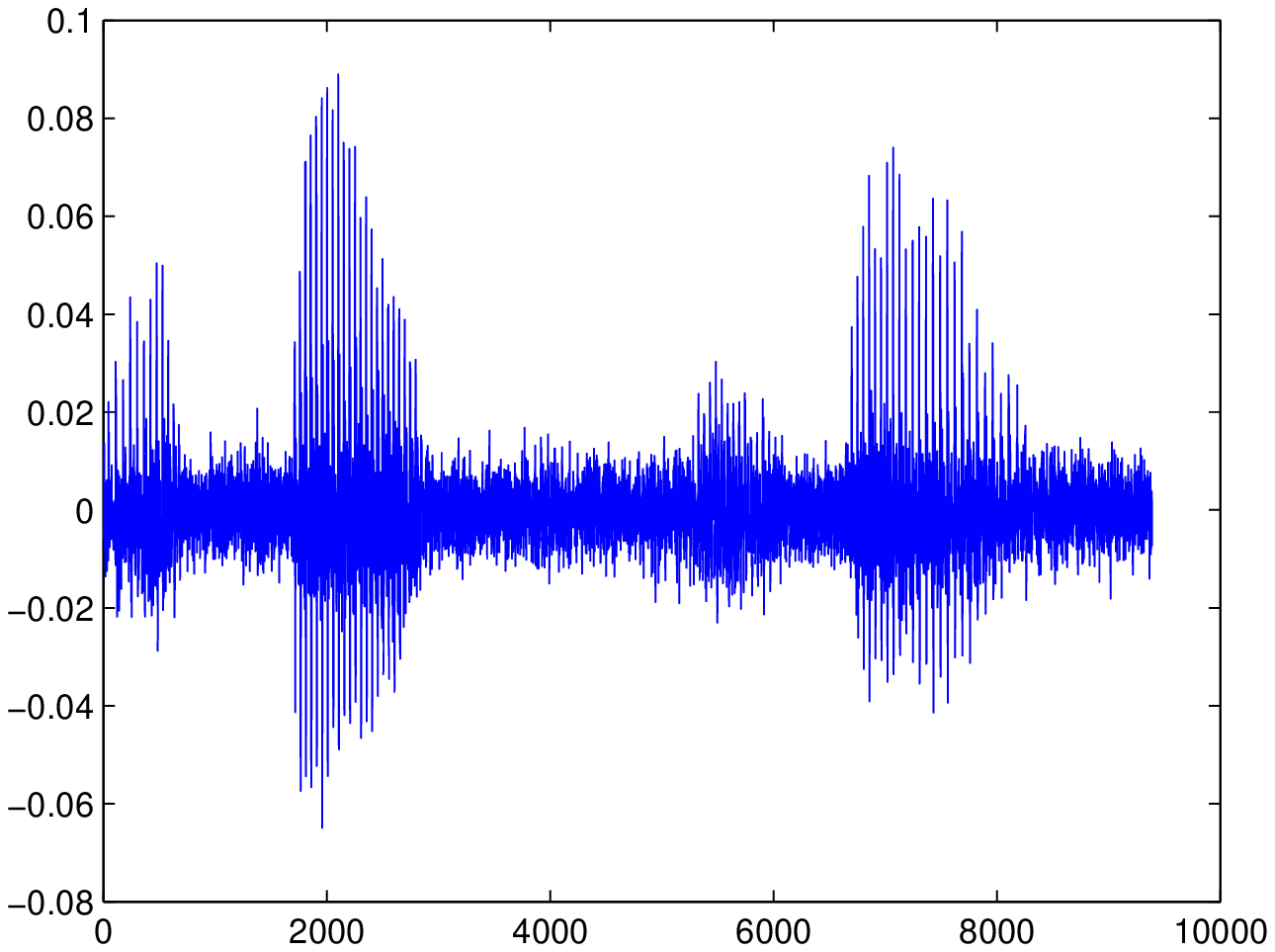}
\caption{Noisy measurement of SPEECH7
 with Gaussian white noise 
and scaled SNR of about $11.1$db.}
\end{figure}
\begin{figure}
\includegraphics[angle= 0, width=0.25\textwidth]{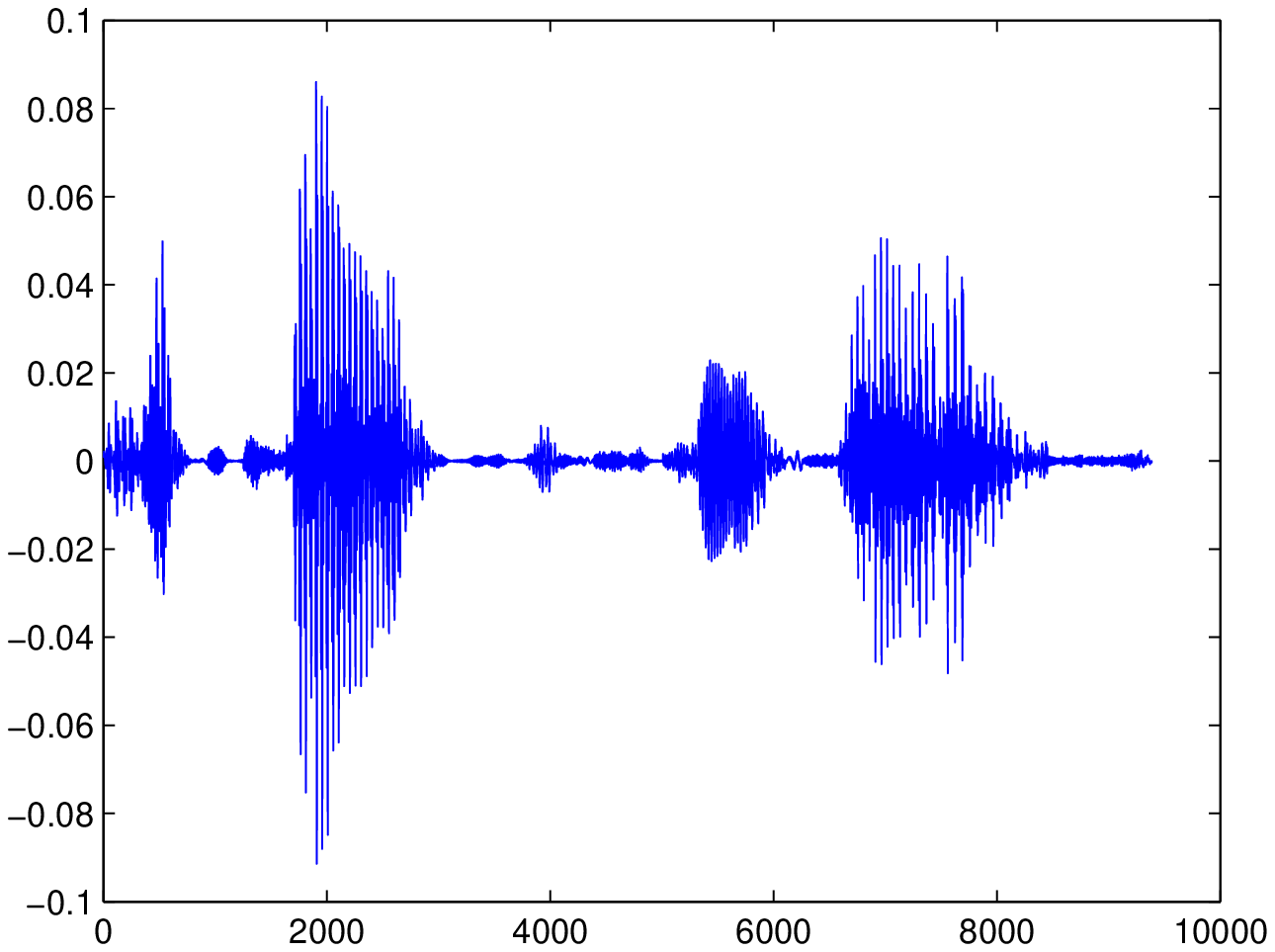}
\caption{Attenuated embedding estimate of SPEECH7 
 from the measurement in Figure 19, scaled to have norm 1,$SNR_s$ is $\approx 7.7$.}
\end{figure}
\begin{figure}
\includegraphics[angle= 0, width=0.25\textwidth]{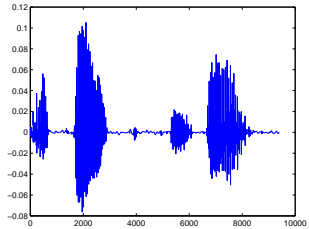}
\caption{Block thresholding estimate of SPEECH7 
 from the measurement in Figure 19, scaled to have norm 1,$SNR_s$ is $\approx 7.6$, note 
low intensity details are removed by the estimator.}
\end{figure}
Data files for the signal, measurement and reconstructions used to compute
the quantities in all the figures   
are available upon request for direct evaluation of the 
perceptual quality.

\section{Further Developments}

Given that the embedding threshold ideas were implemented with the specific
goal of denoising speech signals, it may be worth emphasizing that 
in principle the  construction of classes of paths can be applied to other 
dictionaries well adapted to other classes of signals, more paricularly,
let $\mathcal D=\{g_1,...,g_P\}$ be a generic frame dictionary of $P>N$ elements so that 
$X=\sum_{m=1}^P X_{\mathcal D}[m]\tilde g_m$, $X_{\mathcal D}[m]=<X,g_m>$, 
where $\tilde g_m$ are dual frame vectors
 (see [M] ch.5).
Given such a general representation for $X$,  let 
$\mathcal C_p=\{\gamma_1,...,\gamma_Q\}$, $Q>P$,  be a collection of ordered 
subsets of 
$\mathcal D$
of length $p$, that is, $\gamma_i=\{g_{i_1},...,g_{i_p}\}$, 
so that $\bigcup \gamma_i=\mathcal D$ 
and the cardinality of the set $\{\gamma_i \,\,\text{ such that} \,\, g_j\in \gamma_i\}$ 
is constant for every 
$j=0,...,P-1$ (this ensures that the discrete covering of the frame atoms is 
locally uniform). 
Note that $\mathcal C_p$ needs not be the entire set of 
ordered subsets of
$\mathcal D$. We call each $\gamma_i$ a `path' in $\mathcal D$ for reasons that will be clear in the 
following. Let $X_{\gamma_i}=\{X_{\mathcal D}[m]=<X,g_m>, \,\,g_m \in \gamma_i\}$ be an ordered collection of coefficients 
of $X$ 
in the dictionary $\mathcal D$. 
\\
\\
Then a  a semi-local estimator in $\mathcal D$ can be defined as:
\begin{equation}
\tilde F=\sum_{m=0}^{P-1} d_{I,T}(X_{\mathcal D}[m])\tilde g_m
\label{eqno1} 
\end{equation}
where $d_{I,T}(X_{\mathcal D}[m])=X_{\mathcal D}[m]$ if $I(X_{\gamma})\geq T$ 
for some $\gamma$ 
containing $m$,
and $d_{I,T}(X_{\mathcal D}[m])=0$ if $I(X_{\gamma})< T$ for all 
$\gamma$ containing $m$.
\\
\\
\\
\\
The construction of significant sets of paths $\mathcal C_p$ 
will depend from the application, we are currently exploring even 
the possibility of using 
random walks along  the atoms of the dictionary $\mathcal D$. 
In any case, after $C_p$ is selected,
our specifc choice of index $I^{svd}$ can be used and 
the attenuated embedding estimator can certainly be applied and
tested, soft threshold embedding estimators  are an interesting 
open possibility as well.

\section*{ References}
\begin{description}
\item[{[ABS]}] A. Antoniadis, J. Bigot,  T. Sapatinas,  Wavelet Estimators in Nonparametric Regression: A Comparative Simulation Study, 2001, available http://www.jstatsoft.org/v06/i06/
\item[{[ASY]}]K. T. Alligood, T. D. Sauer, J. A. Yorke, {\it Chaos. An introduction to Dynamical
systems}, Springer, 1996.
\item[{[C]}]T. Cai, Adaptive wavelet estimation: a block thresholding and oracle inequality approach. 
The Annals of Statistics 27 (1999),  898-924.
\item[{[CL]}]T. Cai, M.  Low, 
Nonparametric function estimation over shrinking neighborhoods: Superefficiency and adaptation. 
The Annals of Statistics 33 (2005)., in press. 
 \item[{[CS]}]T. Cai, B. W. Silverman, Incorporating information on neighboring 
coefficients into wavelet estimation, {\it Sankhya} {\bf 63} (2001), 127-148. 
\item[{[DMA]}] G. Davis, S. Mallat and M. Avelaneda, Adaptive Greedy Approximations, Jour. of Constructive Approximation, vol. 13, No. 1, pp. 57-98, 1997
\item[{[DJ]}]D. Donoho, I. Johnstone, Minimax estimation via wavelet shrinkage. {\it Annals of Statistics}{\bf 26} : 879-921,1998.
\item[{[ELPT]}] S. Efromovich, J. Lakey, M.C.  Pereyra, N. Tymes,  
Data-driven and optimal denoising of a signal and recovery of its derivative using multiwavelets,
{\it IEEE transaction on Signal Processing}, {\bf 52} (2004) ,628-635.
\item[{[KS]}]H. Kantz, TSchreiber ,{\it Nonlinear Time Series Analysis}, Cambridge University Press, 2003.
\item[{[HTF]}]T. Hastie, R. Tibshirani, J. Friedman, {\it The Elements of Statistical
Learning}, Springer, 2001.
\item[{[LE]}]E. N. Lorenz, K. A. Emanuel, Optimal Sites for 
Supplementary Weather Observations: Simulation with a Small Model. 
{\it Journal of the Atmospheric Sciences} {\bf 55}, 3 (1998), 399–414.
\item[{[M]}]S. Mallat, {\it A Wavelet Tour of Signal Processing}, Academic Press, 1998.
\item[{[Me]}]A. Mees (Ed.), {\it Nonlinear Dynamics and Statistics}, Birkhauser, Boston, 2001.
\item[{[S]}]T. Strohmer, Numerical Algorithms for Discrete Gabor Expansions, in {\it 
Gabor Analysis and Algorithms. Theory and Applications}, H. G. Feichtinger, T. Strohmer editors.
Birkhauser, 1998.
\item[{[SYC]}]T. Sauer, J. A. Yorke, M. Casdagli, Embedology, {\it Journal of Statistical
 Physics},{\bf 65} (1991), 579-616.

\end{description}
\end {document}